# Prediction of Extraordinary Magnetoresistance in Janus Monolayer MoTeB$_2$

Shijun Yuan,[1,2] Hui Ding,[1] Jinlan Wang[1,*] and Zhongfang Chen[2,*]

*1 School of Physics, Southeast University, Nanjing 211189, P. R. China*
*2 Department of Chemistry, University of Puerto Rico, Rio Piedras Campus, San Juan, PR 00931, USA*

*Corresponding Author: Jinlan Wang (jlwang@seu.edu.cn), Zhongfang Chen (zhongfangchen@gmail.com)

## Abstract

Based on first-principles calculations, we studied the geometric configuration, stability and electronic structure of the two-dimensional Janus MoTeB$_2$. The MoTeB$_2$ monolayer is semimetal, and its attractive electronic structure reveals the perfect electron-hole compensation. Moreover, the electron-type and hole-type bands of the MoTeB$_2$ monolayer are easily adjustable by external stain and charge doping, such as the switch of carrier polarity by charge doping, and the metal-semiconductor transition under tensile stain. These properties allow the MoTeB$_2$ monolayer to be a controllable two-dimensional material with extraordinary large magnetoresistance in magnetic field.

## I. INTRODUCTION

Compared with giant magnetoresistance and colossal magnetoresistance effect, ordinary magnetoresistance (OMR) is usually a relatively weak effect [1]. The OMR effect of nonmagnetic metal is typically reported as the level of one percentage, quadratic only in low field, and tends to saturate in high field. Recently, an extraordinary large magnetoresistance has been discovered in several semimetal materials, such as $PdCoO_2$, $WTe_2$, $Cd_3As_2$, $NbSb_2$, and $MoTe_2$ [2-6]. Unlike the magnetic mechanism of the giant magnetoresistance and colossal magnetoresistance in magnetic multilayers and manganese-based perovskite oxides, several mechanisms have been proposed to explain large OMR in different nonmagnetic compounds [7]. Based on a classical two-band model, the electron-hole compensation is the determining factor for the non-saturating magnetoresistance. Due to the perfect equal-size electron and hole Fermi pockets [8], the quasi-two-dimensional $WTe_2$ layered crystal displays over 4500-fold OMR in a magnetic field of 14.7 Tesla at 4.5 K, and no saturation of OMR even in 60 Tesla [3]. However, unbalanced electron and hole carriers with ultrahigh mobility can also lead to large OMR, such as $PdCoO_2$, $Cd_3As_2$, $NbSb_2$, and $PtSn_4$ [2, 4, 5, 9]. A typical example is $NbSb_2$, which has a small amount of high-mobility hole carrier and a large amount of low-mobility electron carrier [5]. Although there are multiple possible causes of the extremely large magnetoresistance effect, if a certain material has both a perfect electron-hole compensation and high mobility for two-type carriers, it will undoubtedly have a high and non-saturate magnetoresistance effect.

Many two-dimensional materials with a hexagonal boron lattices were predicted to have ultra-high mobility, such as $TiB_2$, $FeB_2$, and $MoB_4$ [10-12]. Since these materials have a graphene-like planar structure, the Dirac cone with a linear energy dispersion emerges near the Fermi level. Each B atom lacks one electron to constitute $s^2p^2$-hybridization for filling the π bands in the boron honeycomb, while metals can just transfer the appropriate number of electrons to stabilize the hexagonal boron structure. Recently, two-dimensional Janus structures with hetero-surface were experimentally prepared, including Janus graphene [13] and Janus MoSSe [14, 15]. Based on chemical intuition, we designed a two-dimensional Janus material $MoTeB_2$, which can be considered as the top layer of tellurium atoms of 1T- or 2H-$MoTe_2$ substituting by double boron atoms. And the Janus monolayer $MoTeB_2$ sheet can also be regarded that a hexagonal Mo layer and a triangular Te layer are sequentially covered on a hexagonal boron lattice. Because the combination of Mo-Te layers can provide exactly two electrons for hexagonal boron honeycomb per unit cell, the $MoTeB_2$ monolayer may be stable. By first-principles calculation, we found that it was a stable semimetal material with a 1:1 electron-hole carrier ratio and high mobility, so an extremely large magnetoresistance can be expected in $MoTeB_2$ monolayer.

## II. COMPUTATIONAL METHOD

All calculations were carried out using the Vienna ab initio simulation package (VASP) code [16, 17] within the projector augmented-wave (PAW) method [18]. Spin-polarized density functional theory (DFT) with the Perdew−Burke−Ernzerhof exchange-correlation potential [19] was applied. The wave energy cutoff of plane-wave basis sets is 500 eV. A vacuum space of at least 15 Å along the out-of-plane direction was employed so that the interaction between two periodic units can be neglected. The *k*-point sampling with a mesh of $35 \times 35 \times 1$ *k*-point generated by the Monkhorst–Pack scheme was used for a primitive cell, and a $7 \times 7 \times 1$ mesh was used for the $4 \times 4 \times 1$ supercell. The lattice constants and atomic coordinates were fully relaxed until the total energy and force converged to $10^{-7}$ eV and to $10^{-3}$ eV/Å, respectively. The phonon frequencies are calculated with $4 \times 4 \times 1$ supercell by using density functional perturbation theory [20] as implemented in the PHONOPY code [21]. The *ab initio* molecular dynamics (AIMD) simulation under a constant-temperature and volume (NVT) ensemble was performed with the temperature controlled at 600 K for a total simulation time of 20 ps with a time step of 1 fs.

## III. RESULTS AND DISCUSSIONS

Figure 1(a) shows the top and side views of the optimized sandwiched Te-Mo-B structure, with two B atoms, one Te atom and one Mo atom in one unit cell. The B atoms are arranged in a honeycomb lattice, with slight buckling height of 0.02 Å. The Te and Mo atoms are arranged in a trigonal lattice, with Te atoms locating at the top of B atoms and Mo atoms siting above the center of the B hexagons. The B-B bond length in MoTeB$_2$ is 1.81 Å, which is between the B-B bond length in an isolated boron honeycomb (1.67 Å) [22] and that of PS$_B$-type MoB$_4$ (1.85 Å).[23] The Mo-B and Mo-Te bond length is 2.34 Å and 2.74 Å, respectively. These lengths are about the same as the Mo-B bond length of 2.37 Å in $MoB_4$ monolayer [12] and the Mo-Te bond length of 2.71 Å in 2H-$MoTe_2$ monolayer [24]. Figure 1(b) gives the phonon dispersion of $MoTeB_2$ monolayer, in which the absence of any imaginary frequency strongly suggests that the structure is a local minimum in its energy landscape. Furthermore, the optical branch is clearly divided into the six high frequency bands from 350 $cm^{-1}$ to 700 $cm^{-1}$, and the three low frequency bands from 150 $cm^{-1}$ to 200 $cm^{-1}$. By comparison with the phonon spectrum of $MoB_4$ monolayer [12], the high frequency branches should correspond to vibrations of the B hexagons and Mo layer, and the low frequency branches come from the Te and Mo lattices. AIMD simulation was further employed to evaluate the kinetic stability of MoTeB$_2$ structure, and Figure 4(c) depicts the fluctuations in the total energy and temperature. Through 20-ps AIMD at 600 K, a MoTeB$_2$ sheet keeps quite original planarity without any lattice destruction, as shown in Fig. 1(d).

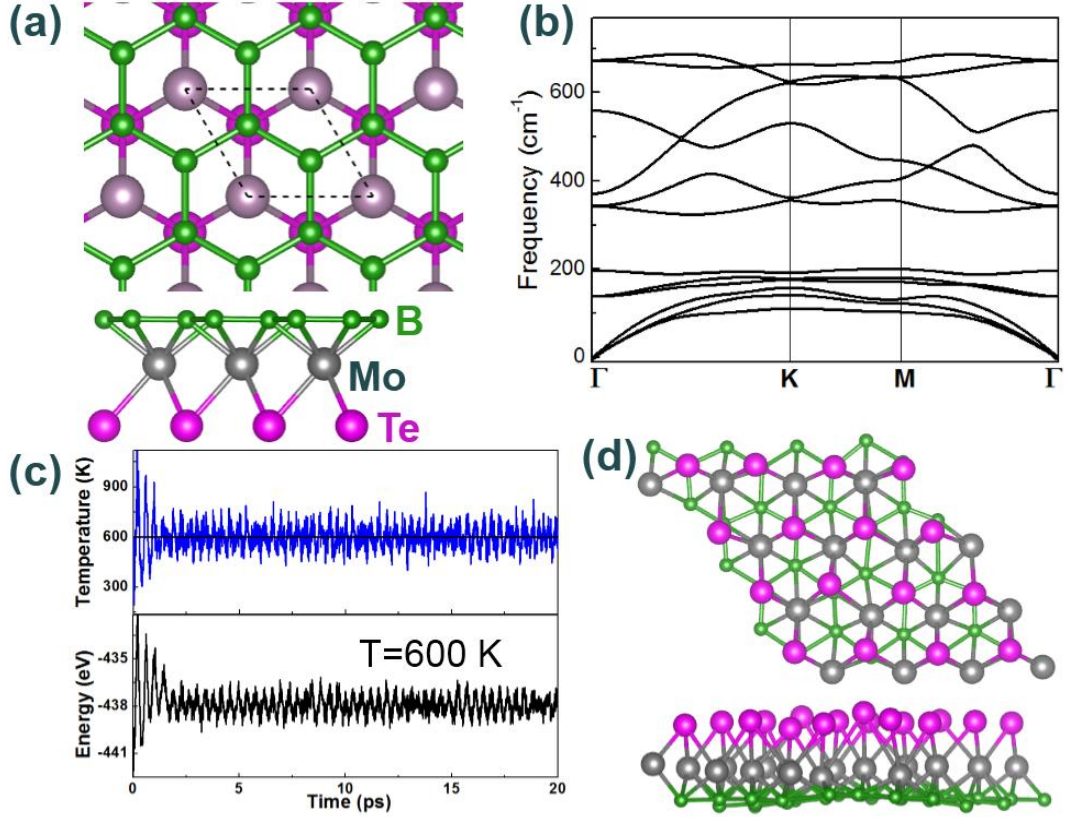

FIG. 1. The structure and its stability of $MoTeB_2$. (a) Top and side views of relaxed structure. (b) The phonon-dispersion curves. (c) The temperature/energy fluctuations depend on simulated time in molecular dynamics simulations at 600 K and (d) the snapshot of a (4 × 4) supercell of $MoTeB_2$ monolayer after a 20ps MD simulation.

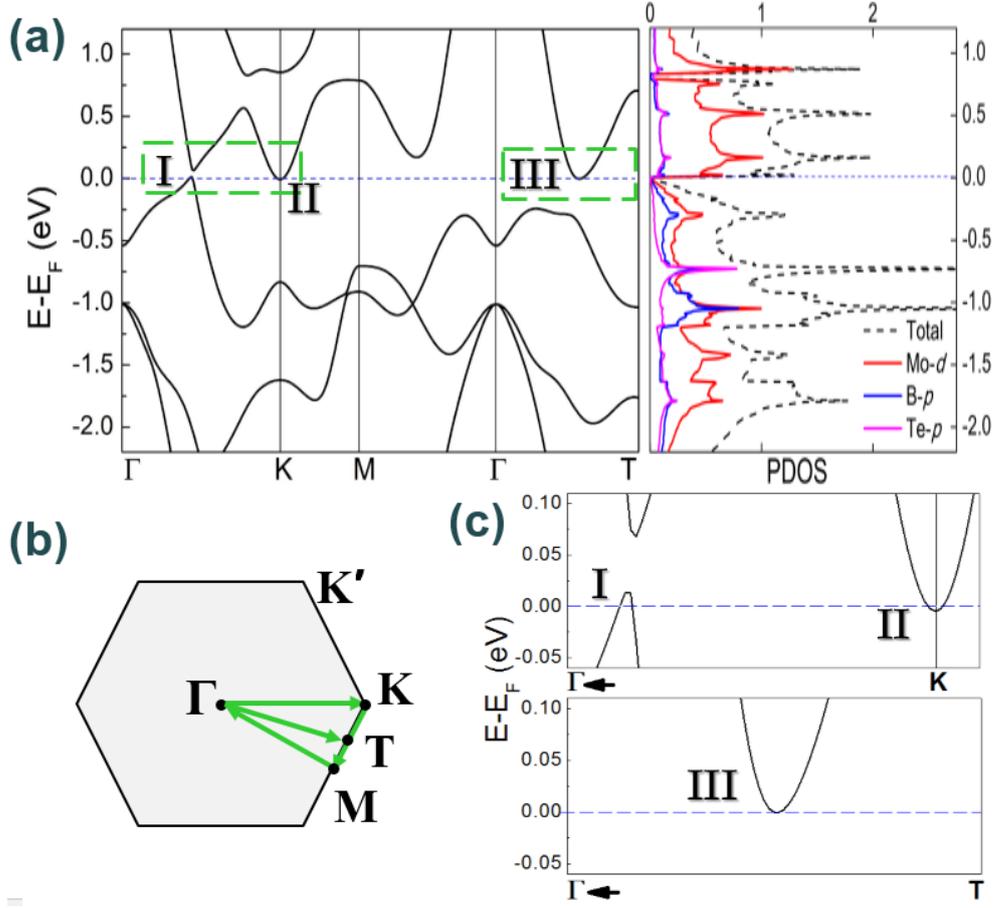

FIG. 2. (a) Electronic band structure and projected density of states (PDOS) for MoTeB$_2$. (b) First Brillouin zone and high-symmetry points along band structure path for MoTeB$_2$. (c) Detail of the calculated electronic structure in the Γ-K and Γ-T directions, which are amplified version of the dashed green boxes labeled in Fig. 2(a). The three intersecting regions between the electronic bands and the Fermi level are marked by I, II, III. Fermi level ($E_F$) is set at the energy zero point.

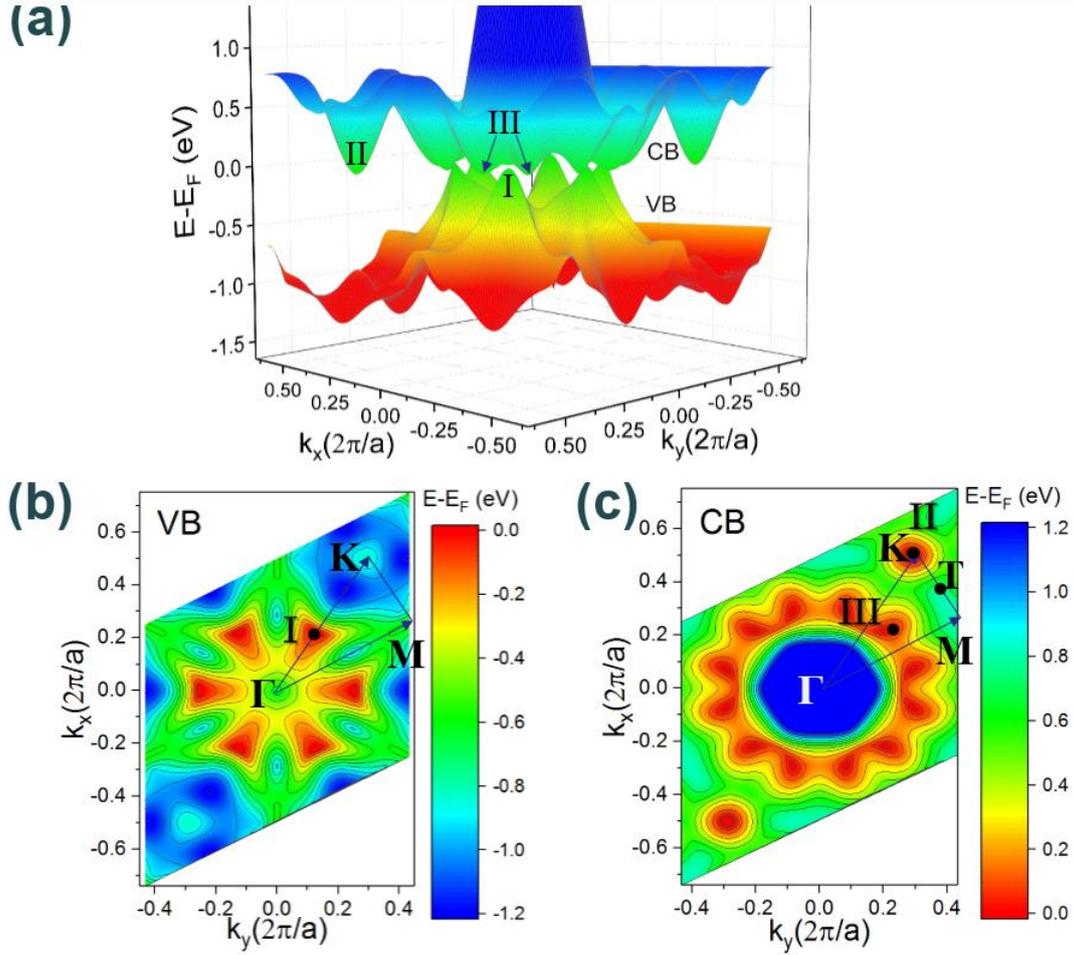

FIG. 3. (a) 3D conduction band (CB) and valence band (VB) of MoTeB$_2$. The constant-energy contours for (b) VB and (c) CB. The black dots at E=0.0 eV labeled as I, II, III, which are the same as in Fig. 2.

The calculated band structures and projected density of states are presented in Fig. 2. The PDOS shows that the states near the Fermi level are mainly contributed by the *d* orbitals of Mo, but the *p* orbitals of B and Te are not negligible. The MoTeB$_2$ monolayer is semimetal, with the valence band (VB) and conduction band (CB) barely crossing the Fermi level at two different places in the Brillouin zone, named by point I and point II in Fig. 2(a). These details can be clearly seen in the enlarged version in Fig. 2(c). The VB and CB almost meet at the point I, with tiny gap of 0.05 eV. The hole-type band around the point I along Γ-K direction is valence band maximum (VBM), whereas the electron-type band around K point (point II) is conduction band minimum (CBM). The CBM is 17 meV below the VBM, which determines the semimetal feature of MoTeB$_2$ monolayer. Meanwhile, there is a contact point between the CB and Fermi level, labeled by the point III in Fig. 2(a). The CB tangent to Fermi energy level (point III) is on the Γ-T line, where the T point is the middle point between the high symmetric points M and K, as shown in Fig. 2(b).

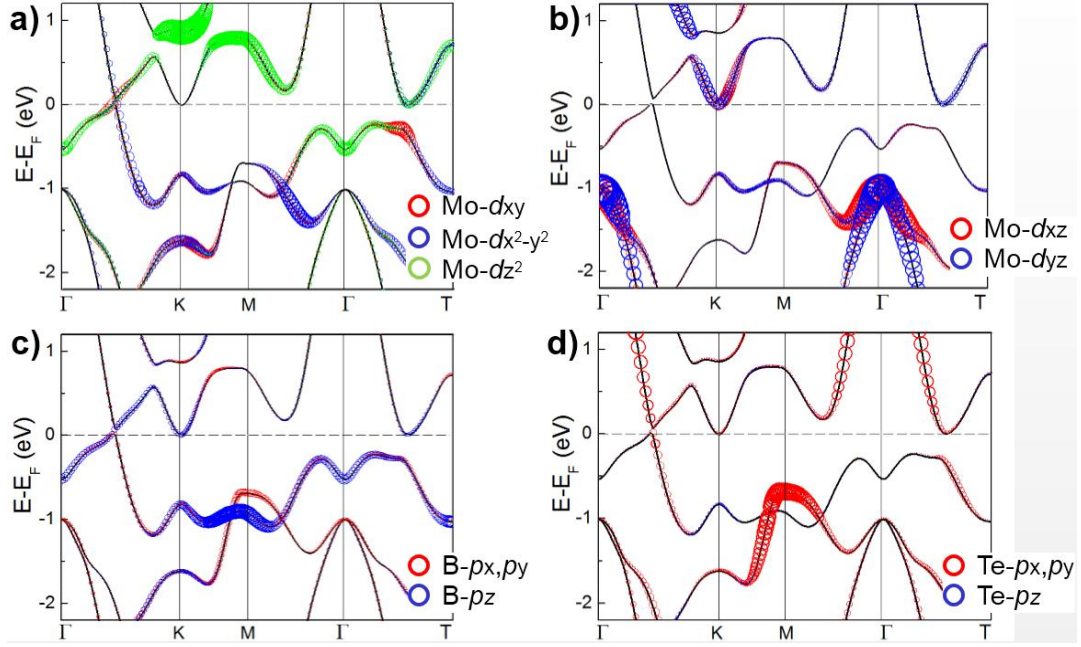

FIG. 4. Projected band structures of (a), (b) $d$ orbitals of Mo atom, $p$ orbitals of (c) B atom and (d) Te atom in MoTeB$_2$, where the radii of the red (blue, green) circles are proportional to the orbital character.

In order to understand the relative position of energy bands near the Fermi level in reciprocal space, we plot the three-dimensional VB and CB in the first Brillouin zone, as shown in Fig. 3(a). The energy peaks of VB (point I) with six-fold symmetry lie right between two energy valleys (point III) of CB. That is, the six VB peaks and the twelve CB valleys occluding like gears, but do not touch each other. Both VB and CB have the same plane $D_6$ symmetry, which is due to the hexagonal lattice symmetry (*p*6m) of MoTeB$_2$ structure.

The transport properties of metal materials are determined by the energy bands passing through the Fermi level, so we focus on the energy bands of MoTeB$_2$ around the point I and point II. As shown in Fig. 3(b), the 2D energy contours of VB and CB exhibit the different morphology of the hole-type band and electron-type band near the Fermi level. The contour lines around VBM (point I) are remarkable anisotropy, and we calculated the slopes of VB in different directions. The slope of VB along the Γ-I direction and the I-K direction in reciprocal space, equal 6.7 eV•Å and -30 eV•Å, respectively. The steep slopes of the VB even can be compared to the linear slope of ±34 eV•Å at the Dirac point in graphene [25]. These direction-dependent slopes are related to the different orbital composition. As shown in Fig. 4(a) and 4(c), the VB exhibits a gentle linear energy dispersion in the Γ-I direction, which are mainly contributed by the hybrid of Mo-$d_{z^2}$ orbitals and B-$p_z$ orbitals. Whereas the linear energy dispersion of VB in the I-K direction is sharp, and this hole-type band are mainly contributed by the hybrid of Mo-$d_{x^2-y^2}$ orbitals and Te-$p_x, p_y$ orbitals, as shown in Fig. 4(a) and 4(d). Since the hybridization between atomic orbitals makes the energy band dispersive, the different slopes of VB along different directions should be due to the stronger interaction between the Mo-Te layers than Mo-B layers. Moreover, the projected band structure around the point I also reveals the origin of the tiny energy gap, which is formed by an avoided hybridization of a Te-$p_x, p_y$ orbital band and B-$p_z$ orbital.

Contrary to the anisotropic hole-type band around VBM, the electron-type band is isotropic around the high symmetric point K, as shown in Fig. 3(c). The quadratic energy band near the Fermi level are mainly contributed by the hybrid of Mo-$d_{xz}, d_{yz}$ orbitals and B-$p_z$ orbitals [Fig. 4(b) and 4(c)], while the contribution of Te atoms is negligible.

Based on the above analysis of the projected bands and 2D energy contours near the Fermi level, it is found that the electron-type carriers and hole-type carriers stem from different sub-bilayer in the Janus MoTeB$_2$ sandwich. The electron-type carriers of CB only come from the hybridization between Mo layer and B layer, while the hole-type carriers of VB come from the hybridization of Mo layer, Te layer and B layer. In Fig. 4, the Mo-Te hybridized band and the Mo-B hybridized band intersect at the point I, which determines the concentration of hole-type carriers.

In Fig. 5, we show an efficient approach based on charge doping to control carrier polarity in MoTeB$_2$ monolayer. The precise position of electron-type band and hole-type band are very sensitive to the Fermi level, so a tiny charge doping can cause a significant energy shift of Fermi level in the MoTeB$_2$ monolayer. After electron or hole doping, the band structures of the MoTeB$_2$ monolayer is still metallic. The electron-doped MoTeB$_2$ monolayer remains only the electron-type Fermi lines and the hole-type Fermi lines vanish. As shown in Fig. 5(a), by the doping of 0.05e per MoTeB$_2$ unit cell, the CB around the point II and point III shifts down below the Fermi level, as showing twelve drop-shape Fermi lines around the point III and one circular Fermi ring around the K point in the first Brillouin zone. In the other hand, under 0.05e hole doping per unit cell, the type carrier of the MoTeB$_2$ monolayer turns to hole-type. As shown in Fig. 5(b), the six trilateral hole-type Fermi lines surround the point I. In the case of a small amount of charge doping, both the linear dispersion of VB and the sharp quadratic dispersion of CB are still maintained. It means that the MoTeB$_2$ monolayer keeps a high hole (electron) mobility regardless of the type of charge injection.

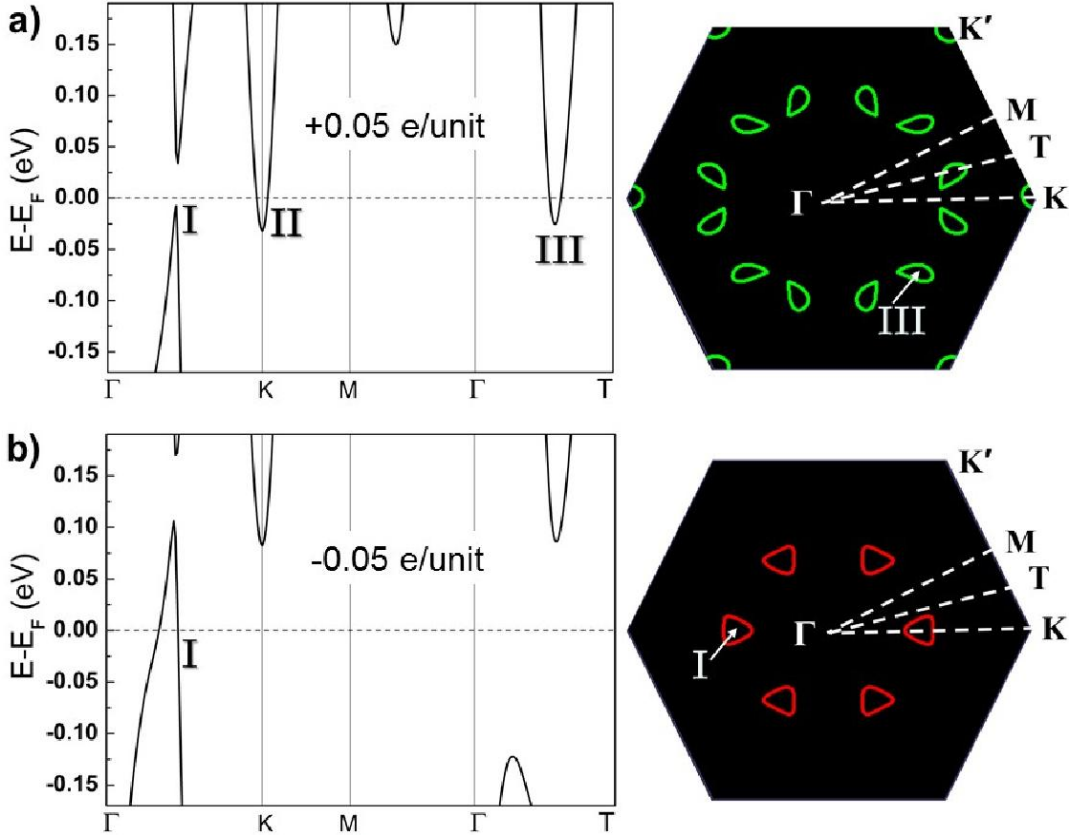

FIG. 5. Band structures and Fermi surface of MoTeB$_2$ doped with electrons or holes. (a) 0.05 electrons doping and (b) 0.05 hole doping per unit cell.

The strain effect on the MoTeB$_2$ monolayer is further investigated because the external strain is effective method to manipulate the electronic properties of low dimensional materials. We found that there was the obvious energy shift of the hole-type and electron-type bands of the MoTeB$_2$ monolayer under isotropic in-plane strains [Fig. 6(a)]. Though the energy band at the point III is insensitive to deformation, which is always tangential to the Fermi level under different strains [Fig. 6(b)], it is noted that the metal-semiconductor transition in the MoTeB$_2$ monolayer under tensile strains. As shown in Fig. 6(c), when the external strain is applied, the hole-type band at the point Γ and the electron-type band at the point K move in the opposite direction. Therefore, the semimetal MoTeB$_2$ monolayer can change to an indirect narrow-bandgap semiconductor by the small tensile strain. On the other hand, in the case of compressive strain from -1% to -3%, the electron-type bands move down rapidly at the point K, while the similar hole-type bands imply the characteristics of VB is independent of compression strains. It indicates that the concentration of the electron-type carriers quickly increases, while the concentration of the hole-type bands remains almost constant under external compressive strain. This property can be used to adjust the electron-hole ratio, and a further result on the electron-hole ratio is shown in Fig. 6(d). The total length of Fermi lines are proportional to the carrier concentration ratio of the $MoTeB_2$ monolayer, so we calculated the perimeters of the electron-type Fermi lines (around point K) and hole-type Fermi lines (around point Γ). By this method, the electron-hole ratio under different external strain can be obtained. For the non-doping $MoTeB_2$ monolayer, the initial electron-hole ratio is 1:1 without external strain. As the compressive strain increases, the election concentration increases faster than the hole concentration, such as the 1.3:1 electron-hole ratio at the 3% compressive strain. Nevertheless, the unbalance electron-hole concentration can be adjusted back to the electron-hole compensation by a slight hole-doping. For example, under 0.01e hole doping per $MoTeB_2$ unit cell, the 1:1 electron-hole ratio occurs at 2% compressive strain. This result means that the carrier concentration of electron and hole are very sensitive to the external strain and change doping, due to the unique electronic structure of $MoTeB_2$ monolayer. Moreover, the specific 1:1 electron-hole ratio can be achieved under various compressive stain, if a small amount of hole-doping is applied on $MoTeB_2$ monolayer. Since the hole-type and electron-type bands of $MoTeB_2$ monolayer exhibit sharp energy dispersion, separation in momentum space, and perfect electron-hole compensation, the $MoTeB_2$ monolayer should be a 2D material with large non-saturating magnetoresistance by proper charge doping and strain.

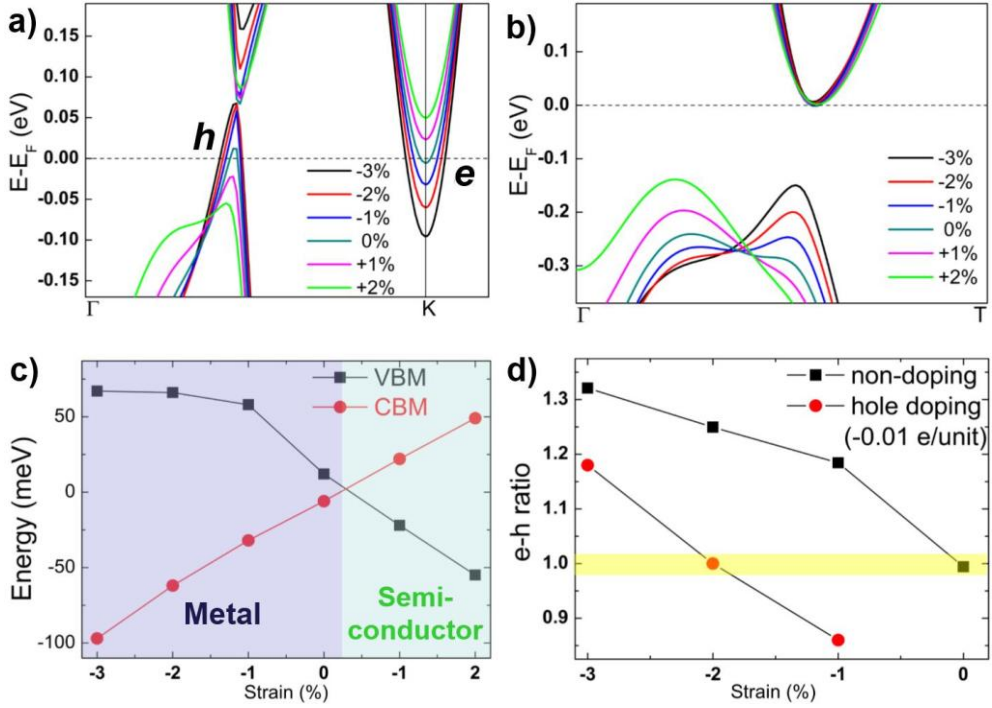

FIG. 6. Band structures of MoTeB$_2$ in the (a) Γ-K and (b) Γ-T directions at different coaxial strains (from -3% compress to +2% stretch). The hole and electron bands are marked with *h* and *e*, respectively. (c) The energy of VBM and CBM under external strains. (d) Strain dependent concentration ratios of electron-type carrier and hole-type carrier without and with hole doping. The yellow area highlights the e-h ratio within 1.00±2%.

**CONCLUSIONS**

In summary, we designed a Janus MoTeB$_2$ monolayer, and systematically examined the geometric structure, stability, electronic properties, strain effect and charge-doped effect by density function theory. The MoTeB$_2$ monolayer has outstanding dynamic and thermal stability, and it is semimetal with CB and VB barely crossing the Fermi level. The electronic property of MoTeB$_2$ monolayer is sensitive to charge doping, so the carrier polarity can switch between electron-type and hole-type by a little charge doping. Moreover, the metal-semiconductor transition occurs in the MoTeB$_2$ monolayer under tensile stain, and the electron-hole carrier ratio of also can be modulated by external strains. These results indicate that the MoTeB$_2$ monolayer has peculiar and adjustable electronic properties, which can be potentially used in magnetoresistance nano-device.

**Acknowledgement**

This work is supported by the National Key R&D Program of China (Grant No. 2017YFA0204800), Natural Science Funds of China (21525311, 21773027), Jiangsu 333 project (BRA2016353). S.J. Yuan gratefully acknowledge financial support from China Scholarship Council.

FIG. 1. The structure and its stability of MoTeB$_2$. (a) Top and side views of relaxed structure. (b) The phonon-dispersion curves. (c) The temperature/energy fluctuations depend on simulated time in molecular dynamics simulations at 600 K and (d) the snapshot of a (4 ×4) supercell of MoTeB$_2$ monolayer after a 20ps MD simulation.

FIG. 2. (a) Electronic band structure and projected density of states (PDOS) for MoTeB$_2$. (b) First Brillouin zone and high-symmetry points along band structure path for MoTeB$_2$. (c) Detail of the calculated electronic structure in the Γ-K and Γ-T directions, which are amplified version of the dashed green boxes labeled in Fig. 2(a). The three intersecting regions between the electronic bands and the Fermi level are marked by I, II, III. Fermi level (E$_F$) is set at the energy zero point.

FIG. 3. (a) 3D conduction band (CB) and valence band (VB) of MoTeB$_2$. The constant-energy contours for (b) VB and (c) CB. The black dots at E=0.0 eV labeled as I, II, III, which are the same as in Fig. 2.

FIG. 4. Projected band structures of (a), (b) *d* orbitals of Mo atom, *p* orbitals of (c) B atom and (d) Te atom in MoTeB$_2$, where the radii of the red (blue, green) circles are proportional to the orbital character.

FIG. 5. Band structures and Fermi surface of MoTeB$_2$ doped with electrons or holes. (a) 0.05 electrons doping and (b) 0.05 hole doping per unit cell.

FIG. 6. Band structures of MoTeB$_2$ in the (a) Γ-K and (b) Γ-T directions at different coaxial strains (from -3% compress to +2% stretch). The hole and electron bands are marked with *h* and *e*, respectively. (c) The energy of VBM and CBM under external strains. (d) Strain dependent concentration ratios of electron-type carrier and hole-type carrier without and with hole doping. The yellow area highlights the e-h ratio within 1.00±2%.